\documentclass{aa}

\usepackage{graphicx}
\usepackage{txfonts}
\usepackage{lscape}
\usepackage{hyperref}
\usepackage{xcolor}
\usepackage{lscape}

\usepackage{pifont}

\newcommand\Halpha{\mbox{${\mathrm H} \alpha$}}%
\newcommand\Hbeta{\mbox{${\mathrm H} \beta$}}%
\newcommand\NII{[\ion{N}{II}]}%
\newcommand{\ngc}[1]{NGC~{#1}}%

\newcommand\nfh{\mbox{$^{\mathrm h}$}}%
\newcommand\nfm{\mbox{$^{\mathrm m}$}}%
\newcommand\solmass{\(\textup{M}_\odot\)}%

\begin{document}

   \title{Discovery of an old nova remnant in the Galactic globular cluster M~22}
   \titlerunning{Discovery of an old nova remnant in M~22}

   \author{Fabian~G\"ottgens \inst{1}
          \and
          Peter~M.~Weilbacher \inst{2}
          \and
          Martin~M.~Roth \inst{2}
          \and
          Stefan~Dreizler \inst{1}
          \and
          Benjamin~Giesers \inst{1}
          \and
          Tim-Oliver~Husser \inst{1}
          \and
          Sebastian~Kamann \inst{3}
          \and 
          Jarle~Brinchmann \inst{4,5}
          \and
          Wolfram~Kollatschny \inst{1}
          \and
          Ana~Monreal-Ibero \inst{7,8}
          \and
          Kasper~B.~Schmidt \inst{2}
          \and
          Martin~Wendt \inst{2,6}
          \and 
          Lutz~Wisotzki \inst{2}
          \and 
          Roland~Bacon \inst{9} 
          }

   \institute{Institut für Astrophysik, Georg-August-Universit\"at G\"ottingen, Friedrich-Hund-Platz 1, 37077 G\"ottingen, Germany\\
              \email{fabian.goettgens@uni-goettingen.de}
         \and 
	     Leibniz-Institut für Astrophysik Potsdam (AIP), An der Sternwarte 16, 14482 Potsdam, Germany 
         \and
             Astrophysics Research Institute, Liverpool John Moores University, 146 Brownlow Hill, Liverpool L3 5RF, United Kingdom 
         \and
          Instituto de Astrof{\'\i}sica e Ci{\^e}ncias do Espaço, Universidade do Porto, CAUP, Rua das Estrelas, PT4150-762 Porto, Portugal 
         \and 
          Leiden Observatory, Leiden University, P.O. Box 9513, 2300 RA, Leiden, The Netherlands
         \and 
	     Institut für Physik und Astronomie, Universität Potsdam, Karl-Liebknecht-Str. 24/25, 14476 Golm, Germany
	 \and    
	     Instituto de Astrof\'{\i}sica de Canarias (IAC), E-38205 La Laguna, Tenerife, Spain
	 \and
	     Universidad de La Laguna, Dpto.\ Astrof\'{\i}sica, E-38206 La Laguna, Tenerife, Spain
	 \and
	     Univ Lyon, Univ Lyon1, Ens de Lyon, CNRS, Centre de Recherche Astrophysique de Lyon UMR5574, F-69230, Saint-Genis-Laval, France
	 }
   \date{Received 6 February 2019; accepted 8 April 2019}

 
  \abstract
   {A nova is a cataclysmic event on the surface of a white dwarf in a binary system that increases the overall brightness by several orders of magnitude. 
   Although binary systems with a white dwarf are expected to be overabundant in globular clusters (GCs) compared to the Galaxy, only two novae from Galactic globular clusters have been observed.
   We present the discovery of an emission nebula in the Galactic globular cluster M~22 (\ngc{6656}) in observations made with the integral-field spectrograph MUSE.
   We extract the spectrum of the nebula and use the radial velocity determined from the emission lines to confirm that the nebula is part of \ngc{6656}.
   Emission-line ratios are used to determine the electron temperature and density.
   It is estimated to have a mass of 1 to $17 \times 10^{-5}$~{\solmass}.
   This mass and the emission-line ratios indicate that the nebula is a nova remnant.
   Its position coincides with the reported location of a `guest star', an ancient Chinese term for transients, observed in May 48 BCE.
   With this discovery, this nova may be one of the oldest confirmed extrasolar events recorded in human history.
   }


   \maketitle
%
\section{Introduction}

   \begin{figure*}
   \centering
      \includegraphics[width=\textwidth]{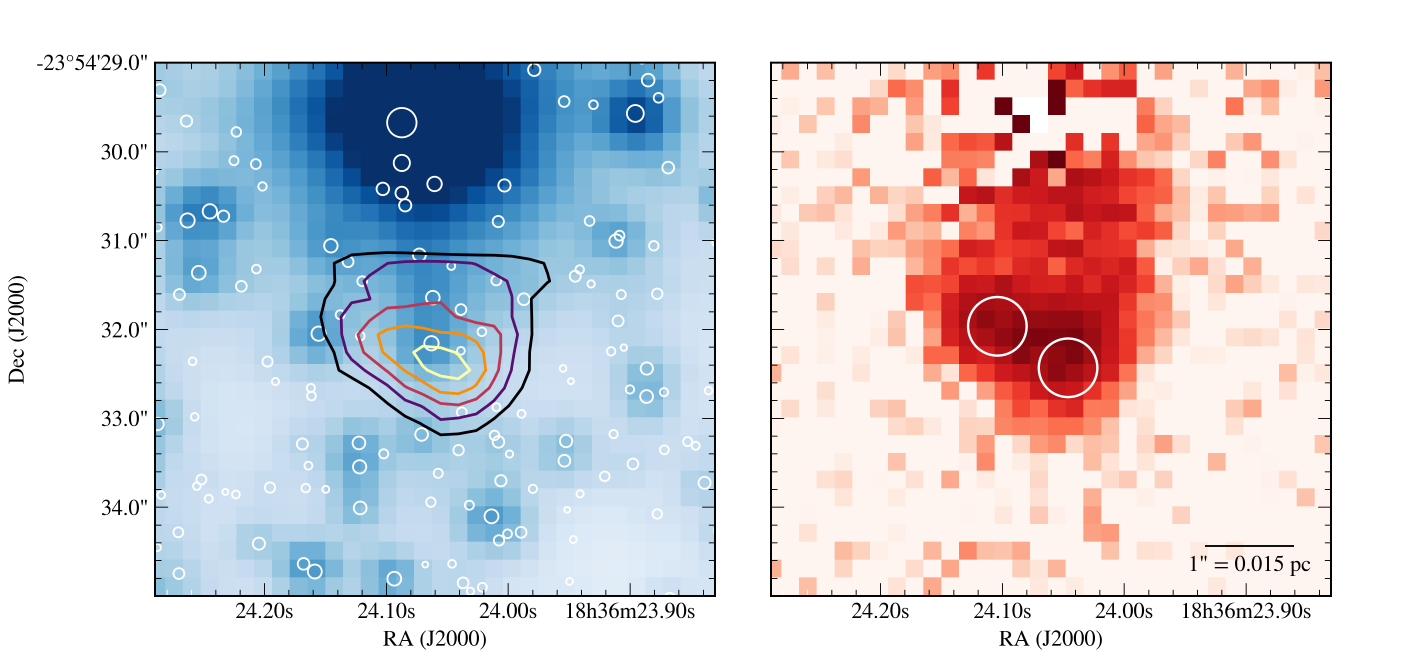}
      \caption{MUSE whitelight detail image of the region in \ngc{6656} containing the nebula created from a single observation (left). 
      Contours give the combined {\NII}$\lambda\lambda$6548,6583 emission flux after subtracting the stellar background (see Section \protect\ref{sct:flux_map}). 
      White circles represent HST sources from the catalogue of \citet{nardiello_hubble_2018}, the diameter scales with F606W magnitude.
      The median flux in the four layers containing the [\ion{N}{II}]$\lambda6583$ emission line is shown after the median flux in three adjacent layers was subtracted (right).
      The two apertures used to extract the spectrum of the nebula are shown as white circles.}
         \label{fig:nebula_map}
   \end{figure*}

Novae are eruptions on the surface of an accreting white dwarf in a cataclysmic variable (CV) binary system \citep{iben_jr._evolution_2008}. 
Hydrogen fusion sets in suddenly when the mass of the accreted hydrogen-rich material on the surface of the white dwarf exceeds a critical value.
The energy set free by fusion causes an eruption on the surface and increases the luminosity by several orders of magnitude. 
The hydrogen-rich matter, possibly mixed with heavier elements of the interior of the white dwarf, is pushed off from the white dwarf with high velocity ($> 10^3$~km/s) and interacts with the interstellar medium.
Although cataclysmic variables are expected to be overabundant in globular clusters compared to the Galactic field \citep{ivanova_formation_2006,knigge_cataclysmic_2012}, 
novae in Galactic globular clusters are observed very rarely.
While there are several observations of novae from extragalactic globular clusters \citep[e. g. ][]{shafter_m31n-2007-06b:_2007,henze_first_2009,henze_supersoft_2013,curtin_exploring_2015-1}, 
there have been only two observations of classical novae (i.~e. a CV without multiple observed eruptions) in Galactic globular clusters: 
T~Sco in the core of \ngc{6093} (M~80) in 1860 \citep{pogson_remarkable_1860,sawyer_bright_1938} and a nova in \ngc{6402} (M~14) in 1938 \citep{hogg_probable_1964}.

Supernovae and novae have been known to Chinese, Arabic, Greek, and Babylonian astronomers for thousands of years \citep{kelley_exploring_2005}.
In Chinese records, these new stars are called `guest stars' because they appear, stay for a while and then disappear.
The oldest Chinese astronomical inscriptions are 3,400 years old and were found on `oracle bones' \citep{pankenier_shang_2015}.
Today, we know that supernovae and novae fit the description of `guest stars', while comets were usually classified differently \citep{stephenson_catalogue_2009}.
For example, the supernova that occurred in 1054 CE was described by several Chinese and Arabic sources \citep{kelley_exploring_2005} and its remnant is known today as the Crab nebula (M~1).
In the case of Nova Scorpii, 1437 CE observed by Korean astronomers, 
\citet{shara_proper-motion_2017} showed that proper motions can be used to identify the CV underlying this nova and to independently determine the age of its remnant.
Even earlier, a `guest star' observed by the Chinese in 77 BCE may have been a dwarf nova outburst of Z Camelopardalis \citep{shara_ancient_2007,johansson_chinese_2007,shara_inter-eruption_2012}, 
although the location of the guest star is very poorly known \citep{stephenson_catalogue_2009}.

Emission nebulae created from ejected material allow observers to investigate the respective supernova or nova that may have happened hundreds or thousands of years ago.
Gas inside globular clusters that could be visible as a nebula is rare as well \citep[see references in][]{barmby_spitzer_2009,lynch_iras_1990}. 
The only visible occurrences in GCs seem to be planetary nebulae (PNe). 
While there are thousands of PNe known in the Milky Way disc, only four PNe have been detected in ${\sim}150$ Galactic globular clusters: 
\object{Ps1} in \ngc{7078} \citep{pease_planetary_1928}, \object{GJJC-1} in \ngc{6656} \citep{gillett_optical/infrared_1989}, \object{JaFu-1} in Pal 6, and \object{JaFu-2} in \ngc{6441} \citep[both][]{jacoby_planetary_1997}.

\ngc{6656} (Messier 22) is one of about 150 Galactic globular clusters, its distance to the Sun is 3.2~kpc \citep[][2010 version]{cudworth_proper_1986,harris_catalog_1996}.
In addition to having a PN, \ngc{6656} sticks out from the set of all Galactic globular clusters because it is one of the few with detected stellar-mass black holes.
\citet{strader_two_2012} detected two accreting stellar-mass black holes in this cluster using X-ray and radio observations which they named M22-VLA1 and -VLA2.
Using numerical models and observational parameters of Galactic GCs, \citet{askar_mocca-survey_2018} predict that \ngc{6656} harbours a population of about 30 stellar mass black holes 
giving rise to its large half-light radius of 1.3~pc \citep[][2010 version]{harris_catalog_1996}.

\section{MUSE observations and data reduction}
\label{sct:data_obs}

\begingroup
\renewcommand{\arraystretch}{1.3}
\begin{table}
\centering
\tiny
\caption{List of MUSE observations of the region containing the nebula. The column \textit{Date} corresponds to the mid-observation time, 
\textit{Seeing} contains the PSF-width measured in the reduced datacubes, and \textit{AO} indicates if the adaptive optics system was used.
}
\begin{tabular}{llll}
\hline
\centering
Date & ESO prog. ID &  Seeing [\arcsec] & AO \\
\hline
2015-05-12 08:25:22 &     095.D-0629 &    0.50 & no \\ 
2015-05-12 08:56:52 &     095.D-0629 &    0.74 & no \\
2015-09-11 02:31:19 &     095.D-0629 &    1.16 & no \\
2015-09-12 02:30:09 &     095.D-0629 &    0.74 & no \\
2016-04-08 09:30:13 &     097.D-0295 &	  0.84 & no \\
2017-04-23 08:06:35 &     099.D-0019 &    0.86 & no \\
2017-04-23 08:45:43 &     099.D-0019 &    0.76 & no \\
2017-10-23 00:52:38 &     099.D-0019 &    0.80 & yes \\
2017-10-23 01:09:09 &     099.D-0019 &    0.74 & yes \\
\hline
\end{tabular}
\label{tab:observations}
\end{table} 
\endgroup

We observed \ngc{6656} during seven nights from 2015 to 2017 with MUSE \citep{bacon_muse_2010}, an integral-field spectrograph at the ESO Very Large Telescope (VLT).
MUSE has a field of view of $1\arcmin \times 1\arcmin$, a spatial sampling of {0\farcs2}, and spectral resolution $R$ between 1800 and 3500 in the spectral range from 4750 to 9350~\AA.
These observations are part of an ongoing survey of 26 Galactic globular clusters 
(PI: S. Dreizler, \citealt{husser_muse_2016,kamann_muse_2016,kamann_stellar_2018,giesers_detached_2018}).
For details on the observations and data reduction, we refer to \citet{kamann_stellar_2018}.
Details about observations of \ngc{6656} are listed in Table~\ref{tab:observations}, including the image quality measured in the final data\-cubes after reduction.
Each MUSE observation of \ngc{6656} has an integration time of 10 minutes.

\section{A new nebula in \ngc{6656}}

As part of a systematic search for emission line sources in Galactic globular clusters (Göttgens et al., in prep.), 
we detected a small emission nebula in \ngc{6656} at a distance of about $14\arcsec$ from the cluster centre.
The region containing the nebula is shown in Fig.~\ref{fig:nebula_map} with a MUSE observation collapsed along the spectral direction 
together with the [\ion{N}{ii}]$\lambda6583$ flux after the flux of adjacent layers is subtracted. 

\subsection{Flux map and spectral properties}
\label{sct:flux_map}

   \begin{figure*}
      \centering
      \includegraphics[width=\textwidth]{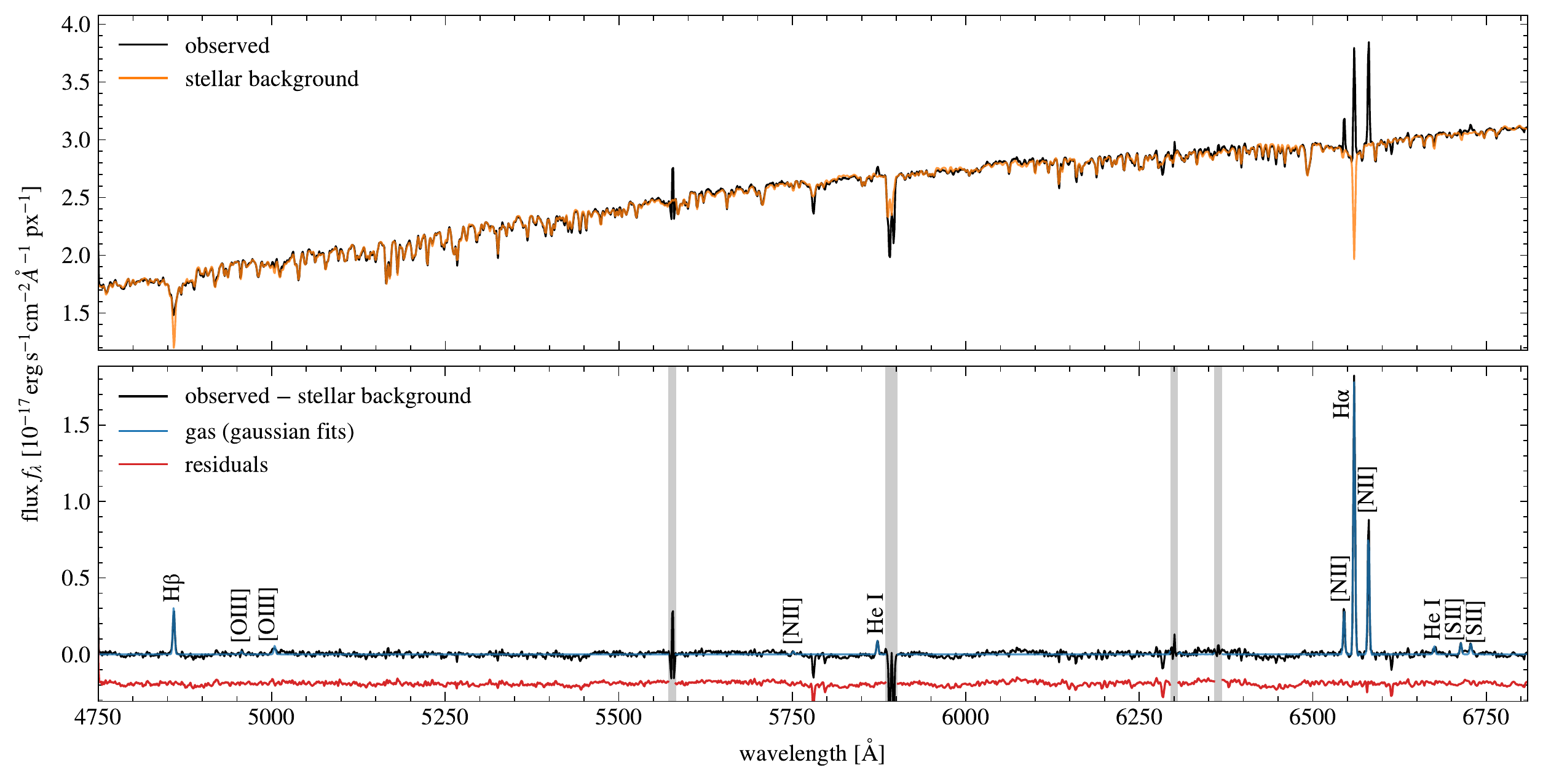}
      \caption{
         Average spectrum extracted from two circular aperatures covering the new nebula (top panel, black). 
         This is decomposed into stellar background (top panel, orange) and gas (bottom panel, black and blue).
	 The residuals after fitting Gaussians to the emission lines are shown in red (bottom panel). 
	 Grey boxes indicate regions dominated by telluric features.
      }
         \label{fig:nebula_spectrum}
   \end{figure*}

Flux maps and spectra extracted with a simple aperture contain a large amount of stellar flux from sources close to the nebula. 
To better disentangle stellar background and ionized gas emission, we employ pPXF \citep[Python version 6.7.12, dated 9 July 2018,][]{cappellari_parametric_2004,cappellari_improving_2017},
the penalized pixel-fitting method that is widely used for full-spectrum fitting with the goal of extracting kinematics of gas and stars and to estimate stellar populations.
We chose to describe the spectra with two sets of templates:

\begin{enumerate}
 \item The stellar background was modelled using the empirical stellar library MILES \citep{sanchez-blazquez_medium-resolution_2006,falcon-barroso_updated_2011}. 
We took the full library of individual stars which samples the expected range of stars in \ngc{6656} well enough. 
However, we did not preselect the stellar spectra, but let pPXF select the best fit. 
 \item The gas emission was modelled with a set of Gaussians, at the positions of the expected relevant lines (see Table \ref{tab:lines}). 
\end{enumerate}
pPXF then optimizes the stellar-background fit and also computes emission-line fluxes. We used 100 Monte-Carlo iterations using the fit residuals to estimate errors of the emission-line fluxes.
As instrumental width, we took the FWHM of the wavelength-dependent MUSE line spread function as computed by the pipeline, convolved to 1.25\,\AA\,pix$^{-1}$ sampling.
As input spectra to pPXF we used two spectra integrated over 0\farcs4 radial apertures (see Fig.~\ref{fig:nebula_map}), placed on the apparent peaks of the {\Halpha} emission,
as well as all individual spectra in the region around the nebula. 
We extracted the spectra from a datacube that combined all available non-AO observing epochs (21 exposures during seven observations with a total integration time of 70 minutes). 
Since the AO data has a slightly different wavelength coverage and a broad gap in the region of NaD and in this case did not actually improve S/N or FWHM significantly, we chose not to include them in the combined
deep dataset.
The contour lines in Fig.~\ref{fig:nebula_map} give the combined {\NII}$\lambda\lambda$6548,6583  emission flux after subtracting the stellar background.
In this map, the nebula appears as an ellipse of $2.5\arcsec \times 2\arcsec,$ corresponding to $0.04\, \text{pc} \times 0.03\, \text{pc}$ 
at the cluster distance of 3.2~kpc \citep{cudworth_proper_1986}.
Fig.~\ref{fig:nebula_map} also shows the [\ion{N}{ii}]$\lambda6583$ spectral layer after subtracting the mean flux of the adjacent layers. 
While this map does not rely on model assumptions, it contains stronger residuals from the bright star above the nebula.
We make the datacube created from all non-AO observations and the extracted spectrum publicly available.\footnote{\url{http://musegc.uni-goettingen.de}}

The average spectrum of the two circular apertures is shown in Fig.~\ref{fig:nebula_spectrum} together with its decomposition into the modelled stellar background and ionised gas.
The spectrum clearly contains strong {\Halpha}, {\Hbeta} and {\NII}$\lambda\lambda$5755,6548,6583 emission lines, as well as weaker emission lines from  [\ion{O}{iii}]$\lambda\lambda$4959,5007, 
[\ion{S}{ii}]$\lambda\lambda$6716,6731, and \ion{He}{i}$\lambda\lambda$5876,6678.

Gaussian fits to the emission lines reveal a line-of-sight (LOS) velocity of $-140 \pm 1$~km/s. 
This is consistent with the assumption that the nebula is comoving with \ngc{6656} which has a LOS velocity of $-146$~km/s \citep[][2010 version]{harris_catalog_1996} 
and a central LOS velocity dispersion of 9~km/s \citep{kamann_stellar_2018}.
The matching LOS velocity and the small apparent separation from the cluster centre of $14\arcsec$ suggest that the nebula is located inside \ngc{6656}.
We further justify this assumption by comparing the expected nova rates from the cluster and the Galactic field (see Appendix).

There are narrowband {\Halpha} HST observations taken with WFPC2 of this region but the nebula is not visible. 

\subsection{Mass of the nebula}
We estimate the mass of the visible nebula using PyNeb \citep[version 1.1.7, dated 18 October 2018,][]{luridiana_pyneb:_2015}
and the equation for the total gas mass given in \citet{corradi_binarity_2015} and used in \citet{sahman_discovery_2018}. 
Since the total amount of intracluster medium in globular clusters is very low \citep[e. g. about 0.3~{\solmass} in the core of \ngc{7078},][]{van_loon_stellar_2006}, we assume that the nebula mass directly corresponds to the mass of the nova ejecta.
This is not true for novae in the Galactic field, as the ejecta sweep up interstellar medium which increases the total mass of the nebula and decelerates its expansion \citep{duerbeck_interaction_1987,shara_proper-motion_2017,darnley_recurrent_2019}.

We use the measured emission line fluxes (given here in units of $10^{-17}\, \text{erg}\, \text{s}^{-1}\, \text{cm}^{-2} $) including 1-$\sigma$ uncertainties of 
[\ion{N}{ii}] $j_{\lambda 5755} = 0.06 \pm 0.08$ and $j_{\lambda 6583} = 1.8 \pm 0.1$ 
together with the 
[\ion{S}{ii}] fluxes of $j_{\lambda 6731} = 0.15 \pm 0.09$ and $j_{\lambda 6716} = 0.13 \pm 0.10$ as input for PyNeb
to estimate an electron temperature 
$T_\mathrm{e} = 1.8 \substack{+1.4 \\ -0.8} \times 10^4 $~K
and an electron density 
$n_\mathrm{e} = 1.2 \substack{+3.5 \\ - 0.9} \times 10^3$~cm$^{-3}.$

To estimate the mass of the nebula, we use the equation given in \citet{corradi_binarity_2015} which requires the total de-reddened {\Hbeta} flux.
We correct the reddening using PyNeb with the extinction law of \citet{fitzpatrick_correcting_1999} and an $E(B-V) = 0.34$ \citep[][2010 version]{harris_catalog_1996}, $R=3.1$, 
and a distance to the nebula of $3.2\pm0.3$~kpc \citep{cudworth_proper_1986}. 
With a total de-reddened {\Hbeta} flux of $(1.9 \pm 0.3) \times 10^{-15}$~$\text{erg}\, \text{s}^{-1}\, \text{cm}^{-2}$, 
we arrive at an estimate for the nebula mass of 1 to $17 \times 10^{-5}$~{\solmass}.
The range takes into account the distance and flux uncertainties as well as the resulting uncertainties in electron temperature and density.

This mass estimate is well inside the expected and observed range for nova shells \citep[see Table 3 in][]{yaron_extended_2005} 
of $10^{-7}$ to $10^{-4}$~{\solmass} (expected) and 1 to $30 \times 10^{-5}$~{\solmass} (observed).

\section{Discussion}

      \begin{figure}
   \centering
      \includegraphics[width=0.5\textwidth]{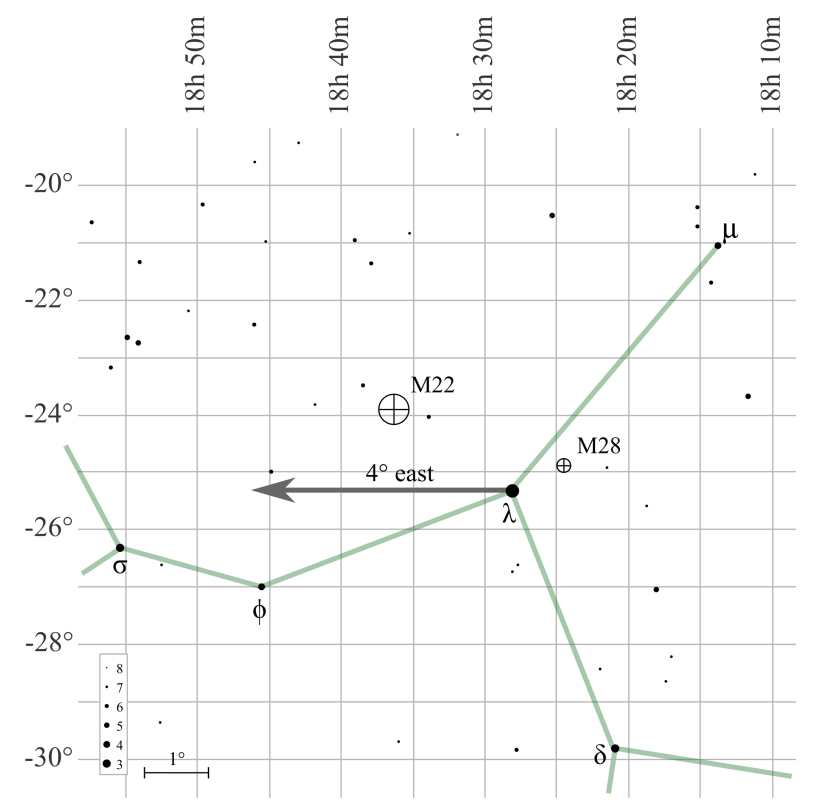}
      \caption{Chart of the sky close to $\lambda$~Sgr as it appeared to a Chinese observer in May 48~BCE.
	       The grey arrow points to the approximate location of the ancient observation.
	       The globular cluster \ngc{6656} (M~22) is located about $2.5^{\circ}$ north-west of this location (about $2.3^{\circ}$ north-east of ${\lambda}$~Sgr).
	       This chart was generated using XEphem \protect\citep[version 3.7.3, ][]{downey_xephem:_2011}.
              }
         \label{fig:skychart}
   \end{figure}

While our mass estimate places the nebula in the range of nova remnants, there are no known novae at this position. 
This raises the question whether there are other indications of a nova origin.

\subsection{Relation to the guest star 48 BCE}

Ancient Chinese records in the \textit{Book of Han} contain a `guest star' observed in May 48 BCE in the Chinese constellation \textit{Nandou} \citep{stephenson_catalogue_2009}. 
The name `guest star' was used by Chinese astronomers for what we today call a nova or a supernova. 
Comets were typically not called `guest stars' -- they had their own category since they could be distinguished by their diffuse appearance and quick apparent motion on the sky.
In the \textit{Book of Han}, the location of the guest star in 48 BCE is given as 4~\textit{chi} east of the second star in \textit{Nandou} (= ${\lambda}$~Sgr), the separation corresponds roughly to 4 degrees~\citep{stephenson_catalogue_2009}.
Fig.~\ref{fig:skychart} illustrates where the stars in the region close to the `guest star' were located in May 48 BCE.
The globular cluster \ngc{6656} is located at ${\rm RA} = 18\nfh 36\nfm 23\fs94, {\rm Dec} = -23^{\circ} 54\arcmin 17\farcs1$, 
about $2.3^{\circ}$ north-east of ${\lambda}$~Sgr in the year 48~BCE, calculated with \texttt{astropy}\footnote{version 3.0.5, \url{http://www.astropy.org}}
using Gaia DR2 coordinates and proper motions \citep{gaia_collaboration_gaia_2018, helmi_gaia_2018}.
According to \citet{stephenson_catalogue_2009}, there is no known supernova remnant within 15~deg of the recorded position.
We argue that the guest star observed in 48 BCE was a nova that occurred in \ngc{6656} and that it is the remnant of this nova that we have detected with MUSE.

\subsubsection{Measurement errors in the year 48 BCE}
The position of the recorded guest star and the observed nova remnant do not coincide exactly. 
\ngc{6656} is rather to the north-east of the reference star instead of east, and the distance is not 4~deg but 2.3~deg.
However, the uncertain conversion of \textit{chi} to degrees and measurement errors in the recorded distance and direction have to be taken into account.
For example, \citet{shara_proper-motion_2017} use a broad range of conversion factors from 0.44 to 2.8 chi/degree.
As determined by \citet{kiyoshi_sogen_1967} and quoted by \citet{ho_peng-yoke_chinese_1972}, the error in stellar positions in observations made about 1000 years later is between 0.5 and 1~deg.
Certainly, this error could not have been smaller in the year 48 BCE.
The Chinese recording associated with the supernova of 1054, which produced the Crab nebula (M~1), even gives a direction to a reference star that is exactly the opposite of what is observed today \citep{ho_peng-yoke_chinese_1972}.
Given these known inaccuracies of ancient Chinese measurements, we are confident that the guest star position matches the position of \ngc{6656}.

\subsubsection{Nova and nebula brightness}
Whether a nova from \ngc{6656} would be visible to the naked eye depends on the distance to the cluster and the absolute nova brightness.
A typical Milky Way nova has an absolute brightness of $-7 \pm 1.4$~mag as determined by \citet{schaefer_distances_2018} using Gaia DR2 parallaxes.
Combined with the distance to \ngc{6656} \citep[3.2~kpc, corresponding to a distance modulus of 12.38~mag,][]{cudworth_proper_1986}, this yields an apparent brightness of $5.38 \pm 1.4$~mag for a GC nova.
Using this estimate, about 40~\% of all novae in \ngc{6656} reach an apparent brightness of 5 mag or brighter, which could be seen with the naked eye.
About 4--5~\% of all novae reach an apparent brightness of at least 3 mag.
This shows that the nova that produced the emission nebula in \ngc{6656} could have been visible to Chinese observers.

Furthermore, we estimate a brightness from the emission line spectrum without the stellar continuum of 25~mag in the Johnson V band.
Using the nova remnant dimming rate of $10 \pm 3$~mmag/year \citep{duerbeck_final_1992} and the nova brightness distribution above,  
we get an age of $2.0 \substack{+0.8 \\ -0.5} \times 10^3$~years which is consistent with the date of the guest star. 

\subsection{Alternative interpretations}
There are several types of emission-line objects that have spectra resembling those of nova remnants, e.g. planetary nebulae.
The mass of a typical Galactic PN is 0.1 to 1 solar masses \citep{osterbrock_astrophysics_1974}, while it can be as low as $10^{-4}$ to $10^{-3}$~{\solmass} for PNe with a binary central star \citep{corradi_binarity_2015}.
Since our estimate yields a mass of 1 to  $17 \times 10^{-5}$~{\solmass} and because of our weak [\ion{O}{iii}]$\lambda5007$ flux, we can exclude a PN as an alternative explanation.
In the case of a PN, one would also expect a very hot and bright (post-AGB) central star as the source of the ionisation energy but such a star is not visible in the HST photometry.
We can exclude that the nebula is a supernova remnant because its flux ratio of [\ion{S}{II}] to {\Halpha} is lower than the canonical value of 0.4 \citep{mathewson_supernova_1973}.
Another explanation would be a merger of two stars which increases the overall brightness and thus could look similar to a nova during outburst. 
The \textit{Nova Vulpeculae 1670} seems to have been such a merger \citep{kaminski_nuclear_2015}, possibly of a white dwarf and a brown dwarf \citep{eyres_alma_2018}.
The bipolar nebula identified with this merger was studied extensively, 
its mass is estimated to be between 0.01 and 0.1 solar masses \citep{eyres_alma_2018} or even as high as 1 solar mass \citep{kaminski_nuclear_2015}.
Thus, we can also exclude a stellar merger as a source of the observed nebula.
Symbiotic stars, i. e. binary systems consisting of interacting red giants and white dwarfs embedded in a nebula fuelled by stellar winds, can have spectra similar to novae. 
We can exclude a symbiotic star as a mimic because there is no red giant star in the centre of the nebula.
We also checked the November 2017 pre-release of the Chandra Source Catalog Release 2.0 \citep{evans_chandra_2010} for X-ray sources close to the nebula that could act as an ionisation source.
The only X-ray source in this region is associated with M22-VLA2, one of the two stellar mass black holes in this globular cluster \citep{strader_two_2012}.

\section{Summary}

We detect a new emission nebula in the globular cluster \ngc{6656} using MUSE integral-field observations. 
After combining exposures from seven observations and modelling the stellar background, we extract a clean spectrum of the nebula.
The spectrum has very strong Balmer and [\ion{N}{ii}] emission lines, as well as several weaker emission lines from [\ion{O}{i}], [\ion{O}{iii}], [\ion{S}{ii}] and \ion{He}{i}.
LOS velocity measurements are consistent with the assumption that the nebula is comoving with \ngc{6656}.
We estimate that the mass of the nebula is between 1 and $17 \times 10^{-5}$~{\solmass}, this estimate is well within the typical observed mass range for the ejecta of classical novae
and outside the typical values for planetary nebulae or stellar merger remnants, which can have a spectrum similar to that of novae.

Ancient Chinese records of astronomical observations include a `guest star', a term used for supernovae and novae, 
which appeared in 48 BCE within ${\sim}2.3$~degrees of the location of \ngc{6656} on the sky. 
The position offset between the recorded guest star and NGC 6656 is within the uncertainty range of ancient observations.
The expected absolute visual brightness of novae at the distance of \ngc{6656} indicates that a cluster nova would have been visible to the naked eye.
We conclude that the nebula detected with MUSE is a nova remnant that was caused by the `guest star' observed roughly 2000 years ago by the Chinese.

\begin{acknowledgements}
We thank the anonymous referee for the constructive report which helped to improve the quality of the work.
We also thank Frederic V. Hessman and Francis R. Stephenson for the helpful discussions.
FG, SK and SD acknowledge support from the German Research Foundation (DFG) through projects KA 4537/2-1 and DR 281/35-1. 
SK gratefully acknowledges funding from a European Research Council consolidator grant (ERC-CoG-646928- Multi-Pop).
PMW, SK, SD and BG also acknowledge support from the German Ministry for Education and Science (BMBF Verbundforschung) through projects MUSE-AO, grants 05A14BAC and 05A14MGA, and MUSE-NFM, grants 05A17MGA and 05A17BAA. 
JB acknowledges support by FCT/MCTES through national funds by this grant UID/FIS/04434/2019 and through the Investigador FCT Contract No. IF/01654/2014/CP1215/CT0003.
AMI acknowledges support from the Spanish MINECO through project AYA2015-68217-P.
RB acknowledges support from the ERC advanced grant 339659-MUSICOS.
Based on observations made with ESO Telescopes at the La Silla Paranal Observatory under programme IDs 094.D-0142, 095.D-0629, 096.D-0175, 097.D-0295, 098.D-0148, 099.D-0019, and 0100.D-0161.
Also based on observations made with the NASA/ESA Hubble Space Telescope, obtained from the data archive at the Space Telescope Science Institute. 
STScI is operated by the Association of Universities for Research in Astronomy, Inc. under NASA contract NAS 5-26555.
\end{acknowledgements}

\bibliographystyle{aa}
\bibliography{nova}

\begin{thebibliography}{58}
\expandafter\ifx\csname natexlab\endcsname\relax\def\natexlab#1{#1}\fi

\bibitem[{Askar {et~al.}(2018)Askar, Arca~Sedda, \&
  Giersz}]{askar_mocca-survey_2018}
Askar, A., Arca~Sedda, M., \& Giersz, M. 2018, Monthly Notices of the Royal
  Astronomical Society, 478, 1844

\bibitem[{Bacon {et~al.}(2010)Bacon, Accardo, Adjali, Anwand, Bauer, Biswas,
  Blaizot, Boudon, Brau-Nogue, Brinchmann, Caillier, Capoani, Carollo, Contini,
  Couderc, Daguisé, Deiries, {B. Delabre}, Dreizler, Dubois, Dupieux, Dupuy,
  Emsellem, Fechner, Fleischmann, François, Gallou, Gharsa, Glindemann, Gojak,
  Guiderdoni, Hansali, Hahn, Jarno, Kelz, Koehler, Kosmalski, Laurent, Floch,
  Lilly, Lizon, Loupias, Manescau, Monstein, Nicklas, Olaya, Pares, Pasquini,
  Pécontal-Rousset, Pello, Petit, Popow, Reiss, Remillieux, Renault, Roth,
  Rupprecht, Serre, Schaye, Soucail, Steinmetz, Streicher, Stuik, H, Vernet,
  Weilbacher, Wisotzki, \& Yerle}]{bacon_muse_2010}
Bacon, R., Accardo, M., Adjali, L., {et~al.} 2010, in Ground-based and
  {Airborne} {Instrumentation} for {Astronomy} {III}, Vol. 7735 (International
  Society for Optics and Photonics), 773508

\bibitem[{Barmby {et~al.}(2009)Barmby, Boyer, Woodward, Gehrz, Loon, Fazio,
  Marengo, \& Polomski}]{barmby_spitzer_2009}
Barmby, P., Boyer, M.~L., Woodward, C.~E., {et~al.} 2009, The Astronomical
  Journal, 137, 207

\bibitem[{Cappellari(2017)}]{cappellari_improving_2017}
Cappellari, M. 2017, Monthly Notices of the Royal Astronomical Society, 466,
  798

\bibitem[{Cappellari \& Emsellem(2004)}]{cappellari_parametric_2004}
Cappellari, M. \& Emsellem, E. 2004, Publications of the Astronomical Society
  of the Pacific, 116, 138

\bibitem[{Corradi {et~al.}(2015)Corradi, García-Rojas, Jones, \&
  Rodríguez-Gil}]{corradi_binarity_2015}
Corradi, R. L.~M., García-Rojas, J., Jones, D., \& Rodríguez-Gil, P. 2015,
  The Astrophysical Journal, 803, 99

\bibitem[{Cudworth(1986)}]{cudworth_proper_1986}
Cudworth, K.~M. 1986, The Astronomical Journal, 92, 348

\bibitem[{Curtin {et~al.}(2015)Curtin, Shafter, Pritchet, Neill, Kundu, \&
  Maccarone}]{curtin_exploring_2015-1}
Curtin, C., Shafter, A.~W., Pritchet, C.~J., {et~al.} 2015, The Astrophysical
  Journal, 811, 34

\bibitem[{Darnley {et~al.}(2019)Darnley, Hounsell, O’Brien, Henze,
  Rodríguez-Gil, Shafter, Shara, Vaytet, Bode, Ciardullo, Davis,
  Galera-Rosillo, Harman, Harvey, Healy, Ness, Ribeiro, \&
  Williams}]{darnley_recurrent_2019}
Darnley, M.~J., Hounsell, R., O’Brien, T.~J., {et~al.} 2019, Nature, 565, 460

\bibitem[{Downey(2011)}]{downey_xephem:_2011}
Downey, E.~C. 2011, Astrophysics Source Code Library, ascl:1112.013

\bibitem[{Duerbeck(1987)}]{duerbeck_interaction_1987}
Duerbeck, H.~W. 1987, Astrophysics and Space Science, 131, 461

\bibitem[{Duerbeck(1992)}]{duerbeck_final_1992}
Duerbeck, H.~W. 1992, Monthly Notices of the Royal Astronomical Society, 258,
  629

\bibitem[{Evans {et~al.}(2010)Evans, Primini, Glotfelty, Anderson, Bonaventura,
  Chen, Davis, Doe, Evans, Fabbiano, Galle, II, Grier, Hain, Hall, Harbo, He,
  Houck, Karovska, Kashyap, Lauer, McCollough, McDowell, Miller, Mitschang,
  Morgan, Mossman, Nichols, Nowak, Plummer, Refsdal, Rots, Siemiginowska,
  Sundheim, Tibbetts, Stone, Winkelman, \& Zografou}]{evans_chandra_2010}
Evans, I.~N., Primini, F.~A., Glotfelty, K.~J., {et~al.} 2010, The
  Astrophysical Journal Supplement Series, 189, 37

\bibitem[{Eyres {et~al.}(2018)Eyres, Evans, Zijlstra, Avison, Gehrz, Hajduk,
  Starrfield, Mohamed, Woodward, \& Wagner}]{eyres_alma_2018}
Eyres, S. P.~S., Evans, A., Zijlstra, A., {et~al.} 2018, Monthly Notices of the
  Royal Astronomical Society, 481, 4931

\bibitem[{Falcón-Barroso {et~al.}(2011)Falcón-Barroso, Sánchez-Blázquez,
  Vazdekis, Ricciardelli, Cardiel, Cenarro, Gorgas, \&
  Peletier}]{falcon-barroso_updated_2011}
Falcón-Barroso, J., Sánchez-Blázquez, P., Vazdekis, A., {et~al.} 2011,
  Astronomy \& Astrophysics, 532, A95

\bibitem[{Fitzpatrick(1999)}]{fitzpatrick_correcting_1999}
Fitzpatrick, E.~L. 1999, Publications of the Astronomical Society of the
  Pacific, 111, 63

\bibitem[{{Gaia Collaboration} {et~al.}(2018){Gaia Collaboration}, Brown,
  Vallenari, Prusti, de~Bruijne, Babusiaux, \&
  Bailer-Jones}]{gaia_collaboration_gaia_2018}
{Gaia Collaboration}, Brown, A. G.~A., Vallenari, A., {et~al.} 2018, Astronomy
  \& Astrophysics, 616, A1, arXiv: 1804.09365

\bibitem[{Giesers {et~al.}(2018)Giesers, Dreizler, Husser, Kamann,
  Anglada~Escudé, Brinchmann, Carollo, Roth, Weilbacher, \&
  Wisotzki}]{giesers_detached_2018}
Giesers, B., Dreizler, S., Husser, T.-O., {et~al.} 2018, Monthly Notices of the
  Royal Astronomical Society: Letters, 475, L15

\bibitem[{Gillett {et~al.}(1989)Gillett, Jacoby, Joyce, Cohen, Neugebauer,
  Soifer, Nakajima, \& Matthews}]{gillett_optical/infrared_1989}
Gillett, F.~C., Jacoby, G.~H., Joyce, R.~R., {et~al.} 1989, The Astrophysical
  Journal, 338, 862

\bibitem[{Harris(1996)}]{harris_catalog_1996}
Harris, W.~E. 1996, The Astronomical Journal, 112, 1487

\bibitem[{Helmi {et~al.}(2018)Helmi, Leeuwen, McMillan, Massari, Antoja, Robin,
  Lindegren, Bastian, Arenou, Babusiaux, Biermann, Breddels, Hobbs, Jordi,
  Pancino, Reylé, Veljanoski, Brown, Vallenari, Prusti, Bruijne, Bailer-Jones,
  Evans, Eyer, Jansen, Klioner, Lammers, Luri, Mignard, Panem, Pourbaix,
  Randich, Sartoretti, Siddiqui, Soubiran, Walton, Cropper, Drimmel, Katz,
  Lattanzi, Bakker, Cacciari, Castañeda, Chaoul, Cheek, Angeli, Fabricius,
  Guerra, Holl, Masana, Messineo, Mowlavi, Nienartowicz, Panuzzo, Portell,
  Riello, Seabroke, Tanga, Thévenin, Gracia-Abril, Comoretto,
  Garcia-Reinaldos, Teyssier, Altmann, Andrae, Audard, Bellas-Velidis, Benson,
  Berthier, Blomme, Burgess, Busso, Carry, Cellino, Clementini, Clotet,
  Creevey, Davidson, Ridder, Delchambre, Dell’Oro, Ducourant,
  Fernández-Hernández, Fouesneau, Frémat, Galluccio, García–Torres,
  González-Núñez, González–Vidal, Gosset, Guy, Halbwachs, Hambly,
  Harrison, Hernández, Hestroffer, Hodgkin, Hutton, Jasniewicz,
  Jean-Antoine-Piccolo, Jordan, Korn, Krone-Martins, Lanzafame, Lebzelter,
  Löffler, Manteiga, Marrese, Martín–Fleitas, Moitinho, Mora, Muinonen,
  Osinde, Pauwels, Petit, Recio-Blanco, Richards, Rimoldini, Sarro, Siopis,
  Smith, Sozzetti, Süveges, Torra, Reeven, Abbas, Aramburu, Accart, Aerts,
  Altavilla, Álvarez, Alvarez, Alves, Anderson, Andrei, Varela, Antiche,
  Arcay, Astraatmadja, Bach, Baker, Balaguer-Núñez, Balm, Barache, Barata,
  Barbato, Barblan, Barklem, Barrado, Barros, Barstow, Muñoz, Bassilana,
  Becciani, Bellazzini, Berihuete, Bertone, Bianchi, Bienaymé,
  Blanco-Cuaresma, Boch, Boeche, Bombrun, Borrachero, Bossini, Bouquillon,
  Bourda, Bragaglia, Bramante, Bressan, Brouillet, Brüsemeister, Brugaletta,
  Bucciarelli, Burlacu, Busonero, Butkevich, Buzzi, Caffau, Cancelliere,
  Cannizzaro, Cantat-Gaudin, Carballo, Carlucci, Carrasco, Casamiquela,
  Castellani, Castro-Ginard, Charlot, Chemin, Chiavassa, Cocozza, Costigan,
  Cowell, Crifo, Crosta, Crowley, Cuypers, Dafonte, Damerdji, Dapergolas,
  David, David, Laverny, Luise, March, Martino, Souza, Torres, Debosscher,
  Pozo, Delbo, Delgado, Delgado, Matteo, Diakite, Diener, Distefano, Dolding,
  Drazinos, Durán, Edvardsson, Enke, Eriksson, Esquej, Bontemps, Fabre,
  Fabrizio, Faigler, Falcão, Casas, Federici, Fedorets, Fernique, Figueras,
  Filippi, Findeisen, Fonti, Fraile, Fraser, Frézouls, Gai, Galleti, Garabato,
  García–Sedano, Garofalo, Garralda, Gavel, Gavras, Gerssen, Geyer,
  Giacobbe, Gilmore, Girona, Giuffrida, Glass, Gomes, Granvik, Gueguen,
  Guerrier, Guiraud, Gutiérrez–Sánchez, Hofmann, Holland, Huckle, Hypki,
  Icardi, Janßen, Fombelle, Jonker, Juhász, Julbe, Karampelas, Kewley, Klar,
  Kochoska, Kohley, Kolenberg, Kontizas, Kontizas, Koposov, Kordopatis,
  Kostrzewa-Rutkowska, Koubsky, Lambert, Lanza, Lasne, Lavigne, Fustec,
  Poncin-Lafitte, Lebreton, Leccia, Leclerc, Lecoeur-Taibi, Lenhardt, Leroux,
  Liao, Licata, Lindstrøm, Lister, Livanou, Lobel, López, Managau, Mann,
  Mantelet, Marchal, Marchant, Marconi, Marinoni, Marschalkó, Marshall,
  Martino, Marton, Mary, Matijevič, Mazeh, Messina, Michalik, Millar, Molina,
  Molinaro, Molnár, Montegriffo, Mor, Morbidelli, Morel, Morris, Mulone,
  Muraveva, Musella, Nelemans, Nicastro, Noval, O’Mullane, Ordénovic,
  Ordóñez–Blanco, Osborne, Pagani, Pagano, Pailler, Palacin, Palaversa,
  Panahi, Pawlak, Piersimoni, Pineau, Plachy, Plum, Poggio, Poujoulet, Prša,
  Pulone, Racero, Ragaini, Rambaux, Ramos-Lerate, Regibo, Riclet, Ripepi, Riva,
  Rivard, Rixon, Roegiers, Roelens, Romero-Gómez, Rowell, Royer, Ruiz-Dern,
  Sadowski, Sellés, Sahlmann, Salgado, Salguero, Sanna, Santana-Ros, Sarasso,
  Savietto, Schultheis, Sciacca, Segol, Segovia, Ségransan, Shih, Siltala,
  Silva, Smart, Smith, Solano, Solitro, Sordo, Nieto, Souchay, Spagna, Spoto,
  Stampa, Steele, Steidelmüller, Stephenson, Stoev, Suess, Surdej, Szabados,
  Szegedi-Elek, Tapiador, Taris, Tauran, Taylor, Teixeira, Terrett,
  Teyssandier, Thuillot, Titarenko, Clotet, Turon, Ulla, Utrilla, Uzzi,
  Vaillant, Valentini, Valette, Elteren, Hemelryck, Leeuwen, Vaschetto,
  Vecchiato, Viala, Vicente, Vogt, Essen, Voss, Votruba, Voutsinas, Walmsley,
  Weiler, Wertz, Wevems, Wyrzykowski, Yoldas, Žerjal, Ziaeepour, Zorec,
  Zschocke, Zucker, Zurbach, \& Zwitter}]{helmi_gaia_2018}
Helmi, A., Leeuwen, F.~v., McMillan, P.~J., {et~al.} 2018, Astronomy \&
  Astrophysics, 616, A12

\bibitem[{Henze {et~al.}(2009)Henze, Pietsch, Haberl, Sala, Quimby, Hernanz,
  Valle, Milne, Williams, Burwitz, Greiner, Stiele, Hartmann, Kong, \&
  Hornoch}]{henze_first_2009}
Henze, M., Pietsch, W., Haberl, F., {et~al.} 2009, Astronomy \& Astrophysics,
  500, 769

\bibitem[{Henze {et~al.}(2013)Henze, Pietsch, Haberl, Valle, Riffeser, Sala,
  Hatzidimitriou, Hofmann, Hartmann, Koppenhoefer, Seitz, Williams, Hornoch,
  Itagaki, Kabashima, Nishiyama, Xing, Lee, Magnier, \&
  Chambers}]{henze_supersoft_2013}
Henze, M., Pietsch, W., Haberl, F., {et~al.} 2013, Astronomy \& Astrophysics,
  549, A120

\bibitem[{{Ho Peng-Yoke} {et~al.}(1972){Ho Peng-Yoke}, Paar, \&
  Parsons}]{ho_peng-yoke_chinese_1972}
{Ho Peng-Yoke}, Paar, F.~W., \& Parsons, P.~W. 1972, Vistas in Astronomy, 13, 1

\bibitem[{Hogg \& Wehlau(1964)}]{hogg_probable_1964}
Hogg, H.~S. \& Wehlau, A. 1964, The Astronomical Journal, 69, 141

\bibitem[{Husser {et~al.}(2016)Husser, Kamann, Dreizler, Wendt, Wulff, Bacon,
  Wisotzki, Brinchmann, Weilbacher, Roth, \& Monreal-Ibero}]{husser_muse_2016}
Husser, T.-O., Kamann, S., Dreizler, S., {et~al.} 2016, Astronomy and
  Astrophysics, 588, A148

\bibitem[{Iben~Jr. \& Fujimoto(2008)}]{iben_jr._evolution_2008}
Iben~Jr., I. \& Fujimoto, M.~Y. 2008, in Classical {Novae}, ed. M.~F. Bode \&
  A.~Evans, Cambridge {Astrophysics} {Series} No.~43 (Cambridge: Cambridge
  University Press)

\bibitem[{Ivanova {et~al.}(2006)Ivanova, Heinke, Rasio, Taam, Belczynski, \&
  Fregeau}]{ivanova_formation_2006}
Ivanova, N., Heinke, C.~O., Rasio, F.~A., {et~al.} 2006, Monthly Notices of the
  Royal Astronomical Society, 372, 1043

\bibitem[{Jacoby {et~al.}(1997)Jacoby, Morse, Fullton, Kwitter, \&
  Henry}]{jacoby_planetary_1997}
Jacoby, G.~H., Morse, J.~A., Fullton, L.~K., Kwitter, K.~B., \& Henry, R. B.~C.
  1997, The Astronomical Journal, 114, 2611

\bibitem[{Johansson(2007)}]{johansson_chinese_2007}
Johansson, G. H.~I. 2007, Nature, 448, 251

\bibitem[{Kamann {et~al.}(2016)Kamann, Husser, Brinchmann, Emsellem,
  Weilbacher, Wisotzki, Wendt, Krajnović, Roth, Bacon, \&
  Dreizler}]{kamann_muse_2016}
Kamann, S., Husser, T.-O., Brinchmann, J., {et~al.} 2016, Astronomy \&
  Astrophysics, 588, A149

\bibitem[{Kamann {et~al.}(2018)Kamann, Husser, Dreizler, Emsellem, Weilbacher,
  Martens, Bacon, den Brok, Giesers, Krajnović, Roth, Wendt, \&
  Wisotzki}]{kamann_stellar_2018}
Kamann, S., Husser, T.-O., Dreizler, S., {et~al.} 2018, Monthly Notices of the
  Royal Astronomical Society, 473, 5591

\bibitem[{Kamiński {et~al.}(2015)Kamiński, Menten, Tylenda, Hajduk, Patel, \&
  Kraus}]{kaminski_nuclear_2015}
Kamiński, T., Menten, K.~M., Tylenda, R., {et~al.} 2015, Nature, 520, 322

\bibitem[{Kelley \& Milone(2005)}]{kelley_exploring_2005}
Kelley, D.~H. \& Milone, E.~F. 2005, Exploring {Ancient} {Skies}: {An}
  {Encyclopedic} {Survey} of {Archaeoastronomy} (New York: Springer-Verlag)

\bibitem[{Kiyoshi(1967)}]{kiyoshi_sogen_1967}
Kiyoshi, Y. 1967, Sōgen jidai no kagaku gijutsu shi (Kyoto: Kyoto Daigaku
  Jimbunkagaku Kenkyusho)

\bibitem[{Knigge(2012)}]{knigge_cataclysmic_2012}
Knigge, C. 2012, Mem.Soc.Ast.It., 83, 549

\bibitem[{Luridiana {et~al.}(2015)Luridiana, Morisset, \&
  Shaw}]{luridiana_pyneb:_2015}
Luridiana, V., Morisset, C., \& Shaw, R.~A. 2015, Astronomy \& Astrophysics,
  573, A42

\bibitem[{Lynch \& Rossano(1990)}]{lynch_iras_1990}
Lynch, D.~K. \& Rossano, G.~S. 1990, The Astronomical Journal, 100, 719

\bibitem[{Marks \& Kroupa(2010)}]{marks_initial_2010}
Marks, M. \& Kroupa, P. 2010, Monthly Notices of the Royal Astronomical
  Society, 406, 2000

\bibitem[{Mathewson \& Clarke(1973)}]{mathewson_supernova_1973}
Mathewson, D.~S. \& Clarke, J.~N. 1973, The Astrophysical Journal, 180, 725

\bibitem[{Nardiello {et~al.}(2018)Nardiello, Libralato, Piotto, Anderson,
  Bellini, Aparicio, Bedin, Cassisi, Granata, King, Lucertini, Marino, Milone,
  Ortolani, Platais, \& van der Marel}]{nardiello_hubble_2018}
Nardiello, D., Libralato, M., Piotto, G., {et~al.} 2018, Monthly Notices of the
  Royal Astronomical Society, 481, 3382

\bibitem[{Osterbrock(1974)}]{osterbrock_astrophysics_1974}
Osterbrock, D.~E. 1974, Astrophysics of {Gaseous} {Nebulae}

\bibitem[{Pankenier(2015)}]{pankenier_shang_2015}
Pankenier, D.~W. 2015, in Handbook of {Archaeoastronomy} and {Ethnoastronomy},
  ed. C.~L. Ruggles (New York, NY: Springer New York), 2069--2077

\bibitem[{Pease(1928)}]{pease_planetary_1928}
Pease, F.~G. 1928, Publications of the Astronomical Society of the Pacific, 40,
  342

\bibitem[{Pogson(1860)}]{pogson_remarkable_1860}
Pogson, N. 1860, Monthly Notices of the Royal Astronomical Society, 21, 32

\bibitem[{Robin {et~al.}(2003)Robin, Reylé, Derrière, \&
  Picaud}]{robin_synthetic_2003}
Robin, A.~C., Reylé, C., Derrière, S., \& Picaud, S. 2003, Astronomy \&
  Astrophysics, 409, 523

\bibitem[{Sahman {et~al.}(2018)Sahman, Dhillon, Littlefair, \&
  Hallinan}]{sahman_discovery_2018}
Sahman, D.~I., Dhillon, V.~S., Littlefair, S.~P., \& Hallinan, G. 2018, Monthly
  Notices of the Royal Astronomical Society, 477, 4483

\bibitem[{Sawyer(1938)}]{sawyer_bright_1938}
Sawyer, H.~B. 1938, Journal of the Royal Astronomical Society of Canada, 32, 69

\bibitem[{Schaefer(2018)}]{schaefer_distances_2018}
Schaefer, B.~E. 2018, Monthly Notices of the Royal Astronomical Society, 481,
  3033

\bibitem[{Shafter \& Quimby(2007)}]{shafter_m31n-2007-06b:_2007}
Shafter, A.~W. \& Quimby, R.~M. 2007, The Astrophysical Journal, 671, L121

\bibitem[{Shara {et~al.}(2017)Shara, Iłkiewicz, Mikołajewska, Pagnotta, Bode,
  Crause, Drozd, Faherty, Fuentes-Morales, Grindlay, Moffat, Pretorius,
  Schmidtobreick, Stephenson, Tappert, \& Zurek}]{shara_proper-motion_2017}
Shara, M.~M., Iłkiewicz, K., Mikołajewska, J., {et~al.} 2017, Nature, 548,
  558

\bibitem[{Shara {et~al.}(2007)Shara, Martin, Seibert, Rich, Salim, Reitzel,
  Schiminovich, Deliyannis, Sarrazine, Kulkarni, Ofek, Brosch, Lépine, Zurek,
  Marco, \& Jacoby}]{shara_ancient_2007}
Shara, M.~M., Martin, C.~D., Seibert, M., {et~al.} 2007, Nature, 446, 159

\bibitem[{Shara {et~al.}(2012)Shara, Mizusawa, Zurek, Martin, Neill, \&
  Seibert}]{shara_inter-eruption_2012}
Shara, M.~M., Mizusawa, T., Zurek, D., {et~al.} 2012, The Astrophysical
  Journal, 756, 107

\bibitem[{Stephenson \& Green(2009)}]{stephenson_catalogue_2009}
Stephenson, F.~R. \& Green, D.~A. 2009, Journal for the History of Astronomy,
  40, 31

\bibitem[{Strader {et~al.}(2012)Strader, Chomiuk, Maccarone, Miller-Jones, \&
  Seth}]{strader_two_2012}
Strader, J., Chomiuk, L., Maccarone, T.~J., Miller-Jones, J. C.~A., \& Seth,
  A.~C. 2012, Nature, 490, 71

\bibitem[{Sánchez-Blázquez {et~al.}(2006)Sánchez-Blázquez, Peletier,
  Jiménez-Vicente, Cardiel, Cenarro, Falcón-Barroso, Gorgas, Selam, \&
  Vazdekis}]{sanchez-blazquez_medium-resolution_2006}
Sánchez-Blázquez, P., Peletier, R.~F., Jiménez-Vicente, J., {et~al.} 2006,
  Monthly Notices of the Royal Astronomical Society, 371, 703

\bibitem[{van Loon {et~al.}(2006)van Loon, Stanimirović, Evans, \&
  Muller}]{van_loon_stellar_2006}
van Loon, J.~T., Stanimirović, S., Evans, A., \& Muller, E. 2006, Monthly
  Notices of the Royal Astronomical Society, 365, 1277

\bibitem[{Yaron {et~al.}(2005)Yaron, Prialnik, Shara, \&
  Kovetz}]{yaron_extended_2005}
Yaron, O., Prialnik, D., Shara, M.~M., \& Kovetz, A. 2005, The Astrophysical
  Journal, 623, 398

\end{thebibliography}

\begin{appendix}
\section{Emission lines used during fitting}
\begingroup
\renewcommand{\arraystretch}{1.3}

\begin{table}[!h]
\centering
\tiny
\caption{List of emission lines that are used to fit Gaussian functions to the observed spectrum.}
\begin{tabular}{ll}
\hline
\centering
Line ID & air wavelength [{\AA}] \\
\hline  
{\Hbeta}    	     &  4861.32 \\
$[\ion{O}{iii}]$4959 &  4958.91 \\
$[\ion{O}{iii}]$5007 &  5006.84 \\
{\ion{He}{i}} 5016      &  5015.68 \\
$[\ion{N}{i}]$5200   &  5199.80 \\ 
$[\ion{N}{ii}]$5755  &  5754.59 \\
{\ion{He}{i}} 5876      &  5875.62 \\
$[\ion{S}{iii}]$6312 &  6312.06 \\
{\Halpha} 	     &  6562.79 \\
$[\ion{N}{ii}]$6548  &  6548.05 \\
$[\ion{N}{ii}]$6583  &  6583.45 \\
{\ion{He}{i}} 6678      &  6678.15 \\
$[\ion{S}{ii}]$6716  &  6716.44 \\ 
$[\ion{S}{ii}]$6731  &  6730.82 \\
  
\hline
\end{tabular}
\label{tab:lines}
\end{table} 
\endgroup

\section{Relative nova rates}
\label{sct:app_inside}
Throughout the analysis we assumed that the nebula is located inside \ngc{6656} as indictated by its matching LOS velocity and its small separation to the centre of \ngc{6656}.
We further justify this assumption by estimating the relative rate of novae originating from the cluster compared to that of the Milky Way stars in the same region of the sky.
The nova rate of a population is the product of its stellar mass and the specific nova rate (i.~e. novae rate per solar mass). 
For the stellar mass, we have to take both foreground and background stars into account because \ngc{6656} is located between the solar system and the Galactic centre at a distance of about 3.2~kpc.
Using a Besan\c con model\footnote{\url{http://model.obs-besancon.fr}, last modified January 18, 2019; used on February 2, 2019} \citep{robin_synthetic_2003} for the Milky Way, we estimate a foreground stellar mass in a  $0.1\deg\,\times\,0.1 \deg$-field centred on \ngc{6656} of $10^3$~{\solmass}
 ($3 \times 10^4$~{\solmass} including background), while the cluster has a mass of $2.9 \times 10^5$~{\solmass} \citep{marks_initial_2010}.
If the specific nova rate is the same for the Milky Way and globular clusters, 
this indicates that there are 300 GC novae per MW nova (or 10 GC novae per MW novae if the total stellar background behind the GC is included). 
The ratio could even be higher if the specific nova rate is higher in globular clusters compared to the Galactic field, as it is the case in M~31 \citep{henze_supersoft_2013}. 
In conclusion, a nova remnant with a small apparent separation to the centre of \ngc{6656} is more likely 
to actually originate from the cluster compared to the Galactic fore- and background due to the higher amount of stellar mass in the cluster.
\end{appendix}

\end{document}